\documentclass[ecta,12pt, a4paper, english, leqno]{article}

\usepackage[left=3.5cm, right=3.5cm, top=3cm, bottom=3cm, includefoot]{geometry} 
\usepackage{sectsty,titlecaps,secdot} 
\sectionfont{\centering \fontsize{12}{15}\selectfont \MakeUppercase} 
\subsectionfont{\centering \fontsize{10}{15}\selectfont \MakeUppercase}
\usepackage{amsmath,amssymb,amsthm,mathtools,stmaryrd,xfrac,dsfont,bm,array,multirow,physics, accents,makecell} 
\newcolumntype{M}[1]{>{\centering\arraybackslash}m{#1}}
\usepackage{natbib} 
\usepackage{graphicx} 
\usepackage[utf8]{inputenc} 
\usepackage{babel} 
\usepackage{float,framed,tikz,tkz-euclide,subcaption} 
\usepackage[shortlabels]{enumitem} 
\usepackage[titletoc]{appendix} 
\usepackage{verbatim} 
\usetikzlibrary{arrows,automata}
\usetikzlibrary{matrix}
\usetikzlibrary{arrows}
\usetikzlibrary{patterns}
\usetikzlibrary{calc}
\usetikzlibrary{shapes.misc}
\usetikzlibrary{positioning,decorations.pathreplacing,shapes}
\usetikzlibrary{decorations.pathreplacing}
\usepackage{pgfplotstable}
\usepackage{pgfplots}
\pgfplotsset{compat=1.12}
\usepgfplotslibrary{fillbetween}
\usepackage[hidelinks]{hyperref} 
\definecolor{Blue}{RGB}{86,180,233}
\hypersetup{colorlinks,linkcolor={Blue},citecolor={Blue}}  

\usepackage{pdfpages} 
\usepackage{pgfplots}
\usepgfplotslibrary{groupplots,colorbrewer}
\pgfplotsset{compat=newest}
\pgfplotsset{cycle list/Set1}
\usepackage{tikz}
\usetikzlibrary{matrix,calc,shapes,arrows.meta,positioning}
\tikzset{
    vertex/.style = {shape=circle,draw, minimum size = 1.8em, inner sep = 0pt},
    edge/.style = {->,> = latex}
}

\usepackage[capitalize,noabbrev]{cleveref}

\newtheoremstyle{break}
{}
{}
{\itshape}
{}
{\bfseries}
{}
{\newline}
{}

\theoremstyle{break}

\newtheorem*{theorem*}{Theorem}
\newtheorem*{cor*}{Corollary}

\newtheorem{prop}{Proposition}
\newtheorem{lem}{Lemma}

\crefname{prop}{Proposition}{Propositions}
\crefname{thm}{Theorem}{Theorems}
\crefname{lem}{Lemma}{Lemmas}
\crefname{blem}{lem}{Lemmas}

\theoremstyle{definition}

\newtheorem*{rem*}{Remark}
\newtheorem*{claim*}{Claim}

\newtheorem*{as*}{Assumption}

\usepackage{accents}
\newcommand{\ubar}[1]{\underaccent{\bar}{#1}}

\def\g{\gamma}

\def\e{\varepsilon}

\def\D{\Delta}

\def\R{\mathbf{R}}

\def\P{\mathbf{P}}

\def\pd{\partial}

\DeclareMathOperator{\E}{\mathbb{E}}

\newcommand{\Paren}[1]{\left( #1 \right)}

\newcommand{\Brac}[1]{\left[ #1 \right]}

\newcommand{\de}{\mathop{}\!\mathrm{d}}

\usepackage{datetime}
\newdateformat{specialdate}{\THEDAY~\monthname[\THEMONTH] \THEYEAR}

\title{On Regularity and Normalization \\ in Sequential Screening}

\date{\specialdate\today}

\author{Ian Ball\thanks{Department of Economics, MIT, ianball@mit.edu} \and Teemu Pekkarinen\thanks{Department of Economics, University of Bonn, pekkariselle@gmail.com}}

\begin{document}

\maketitle

We comment on the regularity assumptions in the multi-agent sequential screening model of \citet[hereafter ES]{esHo2007optimal}. First, we observe that the regularity assumptions are not invariant to relabeling each agent's signal realizations. Second, we show that the regularity assumptions rule out valuation distributions with common bounded support. Third, we show that if each signal realization is labeled to equal the expected valuation, then the regularity assumptions imply that each agent's valuation is equal to his signal realization plus independent mean-zero noise. Fortunately, ES's results go through under the weaker assumption that each agent's virtual value is weakly increasing.

Consider the setting of ES. Each agent $i$ observes a signal $v_i$ about his valuation $V_i$. To simplify notation, we focus on a single agent, and we drop subscripts. The agent's signal $v$ is drawn from a distribution $F$ with continuous density $f$ on $[\ubar{v}, \bar{v}]$, with $f(v) > 0$ for all $v$ in $(\ubar{v}, \bar{v})$. Conditional on the signal realization $v$, the valuation $V$ follows an integrable distribution $H_{v}$ with continuous, strictly positive density $h_{v}$ over $(\ubar{V}, \bar{V})$, where $-\infty \leq \ubar{V} < \bar{V} \leq \infty$.\footnote{The main model of ES assumes that each conditional valuation distribution has positive density over the real line. Many of their examples use restricted supports.} For each fixed $V$ in $(\ubar{V}, \bar{V})$, the map $v \mapsto H_{v} (V)$ is continuously differentiable. There exists an integrable function $b \colon (\ubar{V}, \bar{V}) \to \R_+$ such that $|\pd H_{v} (V) / \pd v| \leq b (V)$ for all $v$ in $[\ubar{v}, \bar{v}]$ and  $V$ in $(\ubar{V}, \bar{V})$.\footnote{This technical assumption allows us to differentiate under the integral.}


In addition to these technical assumptions, we impose one substantive assumption: $\pd H_{v} (V) / \pd v < 0$ for all $v$ in $[\ubar{v}, \bar{v}]$ and $V$ in $(\ubar{V}, \bar{V})$.\footnote{Crucially, we do not require this strict inequality at the endpoints $\ubar{V}$ and $\bar{V}$, since $H_v(\ubar{V}) = 0$ and $H_v (\bar{V}) = 1$ for all $v$. One distribution satisfying our condition is $H_v(V) = V^v$, with $0 < \ubar{v} < \bar{v}$ and $(\ubar{V}, \bar{V}) = (0,1)$. Then $\pd H_v ( V) /\pd v=  (\log V) V^v < 0$ for $V \in (0,1)$. } Thus, higher types have higher valuation distributions, in the sense of the first-order stochastic dominance. 

The agent's signal realization $v$ indexes his conditional valuation distribution. We can relabel the signal realizations, as follows, without changing their informational content. Consider a continuously differentiable bijection $\phi \colon [\ubar{v}, \bar{v}] \to [\ubar{w}, \bar{w}]$ satisfying $\phi'(v) > 0$ for all $v$ in $(\ubar{v}, \bar{v})$. That is, each type $v$ in $[\ubar{v}, \bar{v}]$ is relabeled as $\phi(v)$. The assumptions above are all invariant to such a transformation. 

The assumptions so far are all standing assumptions. Given these standing assumptions, we now analyze the following regularity assumptions from ES.

\begin{enumerate} [label = A\arabic*., ref = A\arabic*] \setcounter{enumi}{-1} 
    \item \label{it:hazard} $f(v) / (1 - F(v))$ is weakly increasing in $v$ over $(\ubar{v}, \bar{v})$. 
    \item \label{it:V_i} $\left( \pd H_{v} (V)/\pd v \right) / h_{v}(V) $ is weakly increasing in $V$ over $(\ubar{V}, \bar{V})$. 
    \item \label{it:v_i} $\left( \pd H_{v} (V)/\pd v \right) / h_{v}(V) $  is weakly increasing in $v$ over $(\ubar{v}, \bar{v})$. 
\end{enumerate}

First, consider how these regularity assumptions are affected by a type transformation $\phi$. This transformation induces new distributions $\tilde{F}$ and $\tilde{H}$ with respective densities $\tilde{f}$ and $\tilde{h}$. For any fixed $w \in [\ubar{w}, \bar{w}]$, let $v = \phi^{-1}(w)$. For all $V$ in $(\ubar{V}, \bar{V})$, we have
\[
    \frac{ \tilde{f} (w)}{1 - \tilde{F}(w)} = \frac{1}{\phi'(v)} \cdot \frac{ f(v)}{1 - F(v)}, 
    \qquad
    \frac{ \pd \tilde{H}_w (V) /\pd w}{\tilde{h}_w (V)} = \frac{1}{\phi'(v)} \cdot \frac{ \pd H_{v} (V)/\pd v}{h_{v}(V)}.
\]
Therefore,
\ref{it:V_i} is invariant to this transformation, but \ref{it:hazard} and \ref{it:v_i} are not. We show that \ref{it:hazard}, in isolation, has essentially no bite. 

\begin{prop}[Type transformation] \label{res:transformation} 
If the inverse hazard rate $(1 - F(v))/f(v)$ is integrable over $[\ubar{v}, \bar{v}]$, then there exists a type transformation $\phi$ under which \ref{it:hazard} holds. 
\end{prop}

In particular, the inverse hazard rate is integrable if the continuous density $f$ is strictly positive on the \emph{closed} interval $[\ubar{v}, \bar{v}]$.\footnote{That is, the density does not converge to $0$ at the endpoints.} In this setting, \ref{it:hazard} has no bite by itself, but the combination of \ref{it:hazard} and \ref{it:v_i} is stronger than \ref{it:v_i} alone. Therefore, it is difficult to interpret \ref{it:hazard} and \ref{it:v_i} separately.

Next, we show that if the common support $[\ubar{V}, \bar{V}]$ of the conditional valuation distributions is bounded below, then the regularity assumptions cannot be satisfied.

\begin{prop}[Impossibility of lower bounded support] \label{res:impossible} If $\ubar{V} > - \infty$, then Assumptions~\ref{it:V_i} and \ref{it:v_i} cannot both hold. 
\end{prop}

Finally, we consider a standard normalization of the signal space. Assumption~\ref{it:hazard} has its usual hazard rate interpretation if signals are  normalized so that $\E [V | v] = v$ for all $v$. Under this normalization, the combination of \ref{it:V_i} and \ref{it:v_i} has a simpler characterization.

\begin{prop}[Normalization implies translation] \label{res:regularity} Suppose $\E [V | v] = v$ for all types $v$.  Assumptions~\ref{it:V_i} and \ref{it:v_i} both hold if and only if $V - v$ is a full-support, mean-zero random variable that is independent of $v$.
\end{prop}

Under the normalization $\E [V | v] = v$, Assumptions~\ref{it:V_i} and \ref{it:v_i} imply that $V = v + \e$ for some full-support, mean-zero random variable that is independent of $v$. (Simply  define $\e = V - v$.) This additive noise specification is natural in many applications. It is not initially apparent, however, that under the standard normalization, the regularity assumptions require this particular specification.

\cref{res:impossible} and \cref{res:regularity} assume that the conditional value distributions have common support. ES's regularity assumptions become more permissive with shifting supports.\footnote{ \cite{LiuEtal2020} show that the much of the analysis goes through with shifting supports. \cite{BP2024} use shifting supports in their analysis of auctions with royalties; that project inspired this note.}
Finally, ES's regularity assumptions \ref{it:hazard}--\ref{it:v_i} are sufficient for their results, but they are not necessary. It is sufficient that the sequential screening virtual value 
\[
    \psi ( v, V) = V +  \frac{1 - F (v)}{f(v)} \frac{ \pd H_{v} (V)/\pd v}{ h_{v}(V)}
\]
is weakly increasing in $v$ and $V$. (With multiple agents, this must hold for each agent's virtual value.)

\appendix
\section{Proofs}

\subsection{Proof of Proposition~1} \label{sec:normaliations}

Define $\phi \colon [\ubar{v}, \bar{v}] \to \R$ by 
\[
    \phi(v) = \ubar{w} + \int_{\ubar{v}}^{v} \frac{1 - F(s)}{f(s)} \de s. 
\] 
This function is well-defined because the hazard rate is integrable. Since $f$ is continuous, $\phi$ is continuously differentiable, with strictly positive derivative on $(\ubar{v}, \bar{v})$. By construction, $\phi'(v) = (1 - F(v))/f(v)$, so the distribution $\tilde{F}$ of $\phi(v)$ has constant hazard rate equal to $1$ and therefore satisfies \ref{it:hazard}.

\subsection{Proof of Proposition~2}

Suppose for a contradiction that $\ubar{V} > -\infty$ and that \ref{it:V_i} and \ref{it:v_i} hold. To simplify notation, let $\g ( v, V) = - ( \pd H_{v} (V)/\pd v) / h_v (V)$. With this notation,  \ref{it:V_i} and \ref{it:v_i} say that $\g$ is weakly \emph{decreasing} in $v$ and $V$. Since $-\pd H_v (V) / \pd v > 0$ for all interior $v$ and $V$, we know that $\g(v, V) > 0$ for all interior $v$ and $V$. Fix $v_0 \in (\ubar{v}, \bar{v})$. Since $\g(v_0, \cdot)$ is decreasing, we have
\[
    \lim_{V \downarrow \ubar{V}} \g( v_0, V) > 0. 
\]
After applying to the type space an increasing affine transformation with sufficiently small positive slope, we may assume that this limit is strictly larger than $1$. So there exists some $V_0$ in $(\ubar{V}, \bar{V})$ such that $\g (v_0, V) >1$ whenever $V \leq V_0$. Since $\g$ is weakly decreasing in $v$, it follows that $\g( v, V) > 1$ whenever $v \leq v_0$ and $V \leq V_0$. 

For the rest of this proof, we write $H( \cdot | v)$ and $h(\cdot | v)$ in place of $H_v$ and $h_v$. Define the function $\D ( v, V) = H( V + v| v)$. Observe that $\D (v, V) = 0$ if $V \leq \ubar{V} - v$ and $\D (v,V) > 0 $ if $V > \ubar{V} - v$. A subscript $j$ on any function denotes the partial derivative with respect to the $j$-th argument. For any $v$ in $(\ubar{v}, v_0)$ and any $V \in (\ubar{V} - v, V_0 - v)$, we have
\begin{align*}
    \D_1 ( v, V)
    &= h (V + v | v) + H_2 ( V + v | v) \\
    &= h(V + v| v) [1- \g(v, V + v)] \\
    &<0.
\end{align*}
Therefore, for any $v_1, v_1', v_2 \in (\ubar{v}, v_0)$ with $v_1 < v_1' < v_2$, we have
\begin{align*}
    H ( \ubar{V} + v_2 - v_1 | v_2) -  H (\ubar{V} + v_1' - v_1 | v_1') 
    &=  \D (v_2, \ubar{V}  - v_1) -  \D(v_1', \ubar{V} - v_1) \\
    &= \int_{v_1'}^{v_2} \D_1 ( v, \ubar{V} -v_1) \de v \\
    &< 0.
\end{align*}
Since $H$ is continuous, passing to the limit as $v_1' \downarrow v_1$, gives 
\[
   H(\ubar{V} + v_2 - v_1 | v_2) \leq \int_{v_1}^{v_2} \D_1 ( v, \ubar{V} -v_1) \de v < 0.
\]
This contradiction completes the proof.

\subsection{Differentiating under the integral}

 Let $\mu (v) = \E [ V | v]$. Write $\P_v$ for the conditional probability measure associated with $H_{v}$. 

\begin{lem}
\label{res:layer} For any type $v$, we have
\[
    \mu'(v) = -\int_{\ubar{V}}^{\bar{V}} (\pd H_v( V) /\pd v) \de V.
\]
\end{lem}

\begin{proof} Since $\mu'$ is invariant to translations of $V$, we may assume without loss that $\ubar{V} < 0 < \bar{V}$. Decompose $V$ into its positive and negative parts, $V_+$ and $V_-$. Apply the layer cake representation to each part separately to get
\begin{equation*}
\begin{aligned}
    \mu(v)
    &=\int_{0}^{\infty} \P_v ( V_+ \geq x) \de x - \int_{0}^{\infty} \P_v ( V_- \geq x) \de x \\
    &=\int_{0}^{\bar{V}} [1 - H_v (x)] \de x - \int_{\ubar{V}}^{0} H_v (x) \de x.
\end{aligned}
\end{equation*}
Since $|\pd H_{v} (V) / \pd v| \leq b (V)$ for all $v$ and $V$, we can differentiate under the integral sign.\footnote{Using the mean-value theorem and the inequality we can show that for any $v$ and any $h > 0$, the difference quotients $h^{-1}\Paren{H_{v + h} (V) - H_v (V)}$ are dominated by the function $b$. Apply Lebesgue's dominated convergence theorem.} Thus,
\[
    \mu'(v) = - \int_{\ubar{V}}^{\bar{V}} \Paren{ \pd H_v(V) / \pd v} \de V. \qedhere
\]
\end{proof}

\subsection{Proof of Proposition~3}
Recall the notation $\g ( v, V) = - ( \pd H_{v} (V)/\pd v) / h_v (V)$ from the proof of Proposition 2. By \cref{res:layer}, since $\mu'(v) = 1$, we have
\begin{equation} \label{eq:one}
    1 = - \int_{\ubar{V}}^{\bar{V}} \Paren{ \pd H_v(V) / \pd v} \de V  = \E_{V | v}  \Brac{ \g (v, V)}.
\end{equation}
Assumptions~\ref{it:V_i} and \ref{it:v_i} say that $\g$ is weakly decreasing in $V$ and $v$. We claim that $\g (v, V) = 1$ for all $v$ and $V$.

First we show that $\g$ is constant in $V$, for each fixed type $v$. Suppose for a contradiction that for some type $\hat{v}$, the function $\g ( \hat{v}, \cdot)$  is not constant. For any type $v$ with $v > \hat{v}$, we have
\begin{equation*}
    \E_{V | v} \Brac{ \g ( v, V)}
    \leq \E_{ V | v}  \Brac{ \g ( \hat{v}, V)} 
    < \E_{V | \hat{v}} \Brac{ \g (\hat{v}, V)},
\end{equation*}
where the weak inequality uses \ref{it:v_i} and the strict inequality holds because $\g (\hat{v}, \cdot)$ is weakly decreasing and not constant, and $H_{\hat{v}} ( V) > H_{v} (V)$ for all $V$.\footnote{Since $\g(\hat{v}, \cdot)$ is not constant, there exist $V_1$ and $V_2$ with $V_1 < V_2$ such that $\g( \hat{v}, V_1) > \g( \hat{v}, V_2)$. Using the layer cake representation, we have
\begin{align*}
    &\E_{ V |\hat{v}}[ \g (\hat{v}, V)] - \E_{V | v} [ \g (\hat{v}, V)]\\
    &= 
    \int_{0}^{\infty} \Brac{ \P_{V| \hat{v}} ( \g (\hat{v}, V) \geq x) - \P_{V| v} ( \g (\hat{v}, V) \geq x)} \de x  \\
    &\geq 
    \int_{\g(\hat{v}, V_2)}^{ \g ( \hat{v}, V_1)} \Brac{ \P_{V| \hat{v}} ( \g (\hat{v}, V) \geq x) - \P_{V| v} ( \g (\hat{v}, V) \geq x)} \de x \\
    &> 0.
\end{align*}
To see that the inequalities hold, observe that $\g (\hat{v}, V) \geq x$ if and only if $V \leq g(x)$ for some weakly decreasing function $g$ that satisfies $V_1 \leq g(x) < V_2$ whenever $\g ( \hat{v}, V_1) \leq x < \g(\hat{v}, V_2)$.}


Having shown that $\g$ is constant in $V$, we know that $\g (v, V) = \bar{\g} (v)$ for some function $\bar{\g}$. But then \eqref{eq:one} implies that $\bar{\g} (v) = 1$ for all $v$, as desired. 

If $\g ( v, V) = 1$ for all $v$ and $V$, then
\begin{equation} \label{eq:density_equality}
   - \frac {\pd H_{v} ( V)}{\pd v} =  h_{v} (V),
\end{equation}
for all $v$ and $V$. Let $\D (v, V) = H_{v} ( V + v)$. We have

\begin{equation*}
    \D_1 (v, V) 
    = - h_{v} (V + v) + h_{v} ( V+ v)
    = 0,
\end{equation*}
where the first term in the derivative is from \eqref{eq:density_equality}. Therefore, the probability $H_{v} ( V + v)$ is independent of $v$. In particular, the translates of the interval $(\ubar{V}, \bar{V})$ agree with $(\ubar{V}, \bar{V})$, so $(\ubar{V}, \bar{V}) = \R$.

\bibliographystyle{aer}
\bibliography{Procurements.bib}

\end{document}